\documentclass[twocolumn]{aastex61}
\pdfoutput=1 %for arXiv submission
\usepackage{amsmath,amstext}
\usepackage[T1]{fontenc}
\usepackage{apjfonts}
\usepackage[figure,figure*]{hypcap}

\shorttitle{Differential Metal Mixing}

\begin{document}

\title{Metal Mixing and Ejection in Dwarf Galaxies is Dependent on Nucleosynthetic Source}

\author{Andrew Emerick}
\affiliation{Department of Astronomy, Columbia University, New York, NY, 10027, USA}
\affiliation{Department of Astrophysics, American Museum of Natural History, New York, NY, USA}
\author{Greg L. Bryan}
\affiliation{Department of Astronomy, Columbia University, New York, NY, 10027, USA}
\affiliation{Center for Computational Astrophysics, Flatiron Institute, 162 5th Ave, New York, NY, 10003, U.S.A}
\author{Mordecai-Mark Mac Low}
\affiliation{Department of Astrophysics, American Museum of Natural History, New York, NY, USA}
\author{Benoit C{\^o}t{\'e}}
\affiliation{Konkoly Observatory, Research Centre for Astronomy and Earth Sciences, Hungarian Academy of Sciences, Konkoly Thege Miklos ut 15-17, H-1121 Budapest, Hungary}
\affiliation{National Superconducting Cyclotron Laboratory, Michigan State University, East Lansing, MI 48824, USA}
\affiliation{Joint Institute for Nuclear Astrophysics -- Center for the Evolution of the Elements, USA}
\author{Kathryn V. Johnston}
\affiliation{Department of Astronomy, Columbia University, New York, NY, 10027, USA}
\author{Brian W. O'Shea}
\affiliation{Department of Computational Mathematics, Science, and Engineering, Michigan State University, East Lansing, MI, 48824, USA}
\affiliation{Department of Physics and Astronomy, Michigan State University, East Lansing, MI, 48824, USA}
\affiliation{National Superconducting Cyclotron Laboratory, Michigan State University, East Lansing, MI, 48824, USA}
\affiliation{Joint Institute for Nuclear Astrophysics -- Center for the Evolution of the Elements, USA}

\begin{abstract}
Using a high resolution simulation of an isolated dwarf galaxy, accounting for multi-channel
stellar feedback and chemical evolution on a star-by-star basis, we investigate how each of 15 metal species are distributed within our multi-phase interstellar medium (ISM) and ejected from our galaxy by galactic winds. For the first time, we demonstrate that the mass fraction probability distribution functions (PDFs) of individual metal species in the ISM are well described by a piecewise log-normal and power-law distribution. The PDF properties vary within each ISM phase. Hot gas is dominated by recent enrichment, with a significant power-law tail to high metal fractions, while cold gas is predominately log-normal. In addition, elements dominated by asymptotic giant branch (AGB) wind enrichment (e.g. N and Ba) mix less efficiently than elements dominated by supernova enrichment (e.g. $\alpha$ elements and Fe). This result is driven by the differences in source energetics and source locations, particularly the higher chance compared to massive stars for AGB stars to eject material into cold gas. Nearly all of the produced metals are ejected from the galaxy (only 4\% are retained), but over 20\% of metals dominated by AGB enrichment are retained. In dwarf galaxies, therefore, elements synthesized predominately through AGB winds should be both overabundant and have a larger spread compared to elements synthesized in either core collapse or Type Ia supernovae. We discuss the observational implications of these results, their potential use in developing improved models of galactic chemical evolution, and their generalization to more massive galaxies.
\end{abstract}


\section{Introduction}
Understanding galactic chemical evolution for all metal species across galaxy mass scales remains one of the most challenging aspects of modeling galaxy evolution. One of the most pressing difficulties is a lack of understanding in exactly how metals propagate from their injection sites from stellar winds or supernovae (SNe) and mix through the phases of the interstellar medium (ISM) into star-forming gas.

One-zone chemical evolution models assume homoegeneous mixing of metals from recent star formation into gas available for future star formation \citep[e.g.][]{Lanfranchi2006b,Kirby2011-III,Cote2017,Andrews2017}. While this assumes that metal mixing in the ISM plays no role in delaying future enrichment in star formation, accounting for this process is challenging. More complex models \citep[e.g.][]{SchonrichBinney2009,Pezzulli2016} employ multiple (usually radial) zones or follow multiple gas phases to attempt to account for these effects. How to properly account for multi-phase mixing, however, is still poorly understood. This is in large part because both these models and large-scale cosmological simulations lack the necessary fidelity to capture the detailed, multi-phase mixing process of metals in the ISM directly. Recent hydrodynamics simulations have employed parametric models to account for unresolved sub-grid metal mixing \citep{PanScannapiecoScalo2013,Sarmento2017,Sarmento2018}, which plays an important role in determining the chemical properties of galaxies \citep[e.g.][]{Shen2010, Pilkington2012, Few2012, Brook2014, FengKrumholz2014, Armillotta2018, Escala2018, Rennehan2018} 
% KVJ: citations to high-redshift work in this area
%\textbf{
and the enrichment process and chemical signatures of the first stars \citep[e.g.][]{Jeon2015,Ritter2015,Smith2015}.
%}
Detailed hydrodynamics simulations incorporating a multi-phase ISM and detailed stellar feedback are required to understand the details of metal mixing and metal outflows in galaxies.

In addition, it remains to be seen to what degree, if at all, metals of different enrichment origins (AGB winds, core collapse SNe, Type Ia SNe, neutron-star neutron-star mergers, etc.) may couple differently to the ISM. When metals are tracked individually, as opposed to a global metallicity field, their injection, mixing, and outflow properties are often treated uniformly. However, if metals do not behave uniformly in the ISM, if their mixing and ejection behavior depends directly upon the energetics and physical environment of their injection, then this assumption would need to be re-evaluated. Differences in metal distributions from ejecta in AGB stars, as compared to SNe, has been explored recently in \cite{KrumholzTing2018}, but has yet to be demonstrated in hydrodynamics simulations. Relaxing this assumption has implications for both interpreting observations of stellar abundances in nearby dwarf galaxies and in modeling galactic chemical evolution in both semi-analytic models and lower resolution cosmological simulations \citep[e.g.][]{Cote2018}.

The use of low mass dwarf galaxies, both observationally in the Local Group and in theoretical models, has been critical in improving our understanding of galactic chemical evolution. That dwarf galaxies efficiently pollute the circumgalactic medium and intergalactic medium (IGM) with metals has been demonstrated for some time both theoretically \citep[e.g.][]{DekelSilk1986,MacLowFerrara1999,Fragile2004,Muratov2017,Corlies2018}, and with direct observational evidence from the metal retention fractions of Local Group dwarf galaxies \citep[e.g.][]{Kirby2011-metals,Bordoloi2014,McQuinn2015}. If there are differences in metal coupling to the ISM, this could potentially impact how metals are driven out of galaxies through galactic winds. This has implications for both interpreting observations of stellar abundances in nearby dwarf galaxies and in modeling galactic chemical evolution in both semi-analytic models and lower resolution cosmological simulations. However, examining this process has only become possible recently as it requires high resolution, galaxy-scale hydrodynamics simulations that can resolve ISM mixing and self-consistently drive galactic winds through multi-channel stellar feedback.
%It is commonly assumed that all metals couple equally to galactic outflows, however, which may not be a valid assumption if metals of different nucleosynthetic origins couple differently to the ISM \citep{KrumholzTing2018}. 
% Relaxing this assumption has implications for both interpreting observations of stellar abundances in nearby dwarf galaxies and in modeling galactic chemical evolution in both semi-analytic models and lower resolution cosmological simulations. However, testing this assumption has only become possible recently as it requires high resolution, galaxy-scale hydrodynamics simulations that can resolve ISM mixing and self-consistently drive galactic winds through multi-channel stellar feedback.

It is becoming increasingly valuable to develop a concrete theoretical understanding of galactic chemical evolution as a result of multiple, recent observational campaigns to probe detailed stellar abundances in our Milky Way and the Local Group, such as SEGUE \citep{Yanny2009}, RAVE \citep{Kunder2017}, APOGEE \citep{Anders2014}, and GALAH \citep{Buder2018}. Stellar abundances are directly imprinted with the enrichment pattern of their star-forming cloud, whose chemical properties are determined by the process of turbulent metal mixing in the ISM. The degree to which we can associate "chemically tagged" stars as co-eval depends directly upon our understanding of metal enrichment and metal mixing in the ISM. Furthermore, this understanding is critical for using these observations to further deduce properties of a galaxy's evolutionary history. 
% KVJ asked to bring up this area of research and emphasize how it is lacking theoretical input
%\textbf{

One key aspect of this work has been a significant effort to characterize the number of independent dimensions accessible by the multi-dimensional chemical abundances observed in these studies \citep[e.g][]{Freeman2002,Ting2012, Hogg2016,Jofre2017,Price-Jones2018}. Thus far, these studies have remained uninformed by hydrodynamics simulations and, conversely, these results cannot yet be used to constrain simulations. Doing so requires the kinds of high resolution, galaxy-scale multi-element chemodynamical simulations that have only become feasible in recent years. Recent work has suggested that low-mass dwarf galaxies are perhaps the best regime to begin understanding these processes \citep[e.g.][]{Bland-Hawthorn2010a,Bland-Hawthorn2010b,Karlsson2012,Webster2016}. This is advantageous, as their small physical size and low star formation rates makes conducting high resolution, hydrodynamics simulations of these systems computationally feasible.

In this paper we present the first detailed chemical evolution results from a set of high-resolution hydrodynamics simulations of an isolated, low-mass, dwarf galaxy performed with the adaptive mesh refinement code \textsc{Enzo} \citep{Enzo2014}. The simulation discussed here was introduced in detail in \cite{Emerick2018} (hereafter Paper I). To address the outstanding questions discussed above, these simulations follow star formation using individual star particles, including stellar feedback from massive star and AGB-phase stellar winds, photoelectric heating, Lyman Werner dissociation, ionizing radiation tracked through an adaptive ray-tracing radiative transfer method, and core collapse and Type Ia SNe. This is in addition to a detailed model for ISM physics using the \textsc{Grackle} library, as discussed below. We show that metals are strongly ejected via galactic winds, but that the retention of metals in the ISM and their mixing through phases varies significantly depending on the production source of the given elemental species. We show how these elements are distributed in the ISM and conclude with a discussion on the implications of these results.

The physical processes that drive galactic chemical evolution are complex, driven by the details of stellar feedback, turbulence and diffusion in a multi-phase ISM, and variations in stellar yields with nucleosynthetic source and stellar metallicity. Uncertainties in each of these processes make reproducing and interpreting observations of gas and stellar abundances challenging. 
%[list examples of modeling problems]. 
These uncertainties, combined with the difficulty in simulating a fully self-consistent galaxy in detail, motivates this current study. By focusing on a low mass dwarf galaxy, with small size and low star formation rate, we can capture detailed feedback and ISM physics at high resolution, while following individual stars. In this work we focus on theoretical quantities, namely the probability density distributions (PDFs) of metals in the ISM as a function of metal mass fraction, rather than the common observables of stellar and gas abundance ratios, for two reasons. First, we would like to build our understanding of galactic chemical evolution from a fundamental level. Second, there is only a limited sample of gas-rich, star forming dwarf galaxies of this size that we can use for direct observational comparison, and it is computationally infeasible to simulate galaxies of easily observable size in as much detail as done here. This work will be the first of several attempting to bridge this gap. With a more fundamental understanding of metal mixing and stellar enrichment in galaxies, we can construct better one-zone chemical evolution models and physically motivated sub-grid physics models for lower resolution simulations of more massive galaxies.

We summarize our methods and physics models in Section~\ref{sec:methods}. We begin our analysis by presenting the only direct observable comparison we can make at this galaxy mass, discussing the metal retention fractions of our galaxy in Section~\ref{sec:ejection}. In Section~\ref{sec:log-normal} we focus on how each of our 15 individual metal species are distributed and evolve within each phase of the ISM. We discuss our results in Section~\ref{sec:discussion}, and conclude in Section~\ref{sec:conclusions}.

% AE: Benoit suggested subsections instead of paragraph splits... play with this to see what looks best
\section{Methods}
\label{sec:methods}
We refer the reader to Paper I for a detailed description of our numerical methods and feedback models. We briefly summarize the relevant details here.

\paragraph{Hydrodynamics} We use the adaptive mesh refinement astrophysical hydrodynamics and N-body code \textsc{Enzo}\footnote{http://enzo-project.org/} \citep{Enzo2014}. Hydrodynamics are solved using a direct-Eulerian piecewise parabolic method and a two-shock approximate Riemann solver with progressive fallback to more diffusive solvers. We include gas self-gravity and evolve collisionless star particles using a particle mesh N-body solver. We use a 128$^{3}$ base grid with outflow boundary conditions measuring 2.16~R$_{vir}$ on a side, where R$_{\rm vir}$ = 27.4~kpc, and 9 levels of refinement, for a maximum spatial resolution of 1.8~pc. Refinement occurs when either: 1) a cell contains more than 50~M$_{\odot}$ of gas or 2) a cell's local Jeans length becomes resolved by less than eight cells. Also, if the cell is within four zones of a star particle with active feedback, it is refined to the maximum resolution. At the maximum resolution, we use a pressure floor to prevent artificial fragmentation when the Jeans length is unresolved.

\paragraph{Chemistry, Heating, and Cooling Physics} We use the the astrophysical chemistry and cooling package \textsc{Grackle} \citep{GrackleMethod} to evolve a nine species non-equilibrium primordial chemistry model (H, H$^+$, He, He$^+$, He$^P{++}$, H$^-$, H$_2$, H$_2^+$, and e$^-$), follow approximate metal line cooling using a \textsc{Cloudy} look-up table, and apply heating from a metagalactic UV background \citep{HM2012}. We account for approximate self-shielding of H~{\sc i} against the UV background following \cite{Rahmati2013}. We assume He~{\sc i} self-shields in the same fashion as H~{\sc i}, and ignore He~{\sc ii} heating from the UVB entirely. Approximate H$_2$ self-shielding from background Lyman-Werner (LW) radiation is accounted for using the Sobolov-like method from \cite{Wolcott-Green2011}. Finally, we use the updated metal line cooling tables which self-consistently account for the decrease in metal line cooling rates due to lower ionization fractions in self-shielding gas, as compared to metal cooling tables computed under the optically thin assumption.

% AE: From Benoit, maybe make a note of 1 - 100 M_sun sampling may slightly overestimate higher mass stars (people usually do 0.1 to 100 M_sun
\paragraph{Star Formation} Stars are followed as individual star particles from 1~M$_{\odot}$ to 100~M$_{\odot}$. Stars are able to form in dense gas with: 1) n~$>$~200~cm$^{-3}$, 2) T~$<$~200~K, and 3) $\nabla \cdot v < 0$. Given the short time-steps ($dt \sim 500$~yr) and high resolution (1.8~pc) in these simulations, the local star formation rate in any single zone is very small ($\ll$ 1~M$_{\odot}$ dt$^{-1}$). We therefore form stars stochastically, depending upon the local gas mass, free-fall time, and star formation efficiency, $\epsilon_{\rm f}$, taken to be 2\% \citep{KrumholzMcKee2005}. Stellar masses are randomly sampled from an assumed \cite{Salpeter1955} IMF with metallicities and metal fractions set by the local gas environment. We use the zero age main sequence properties of stars from the \textsc{PARSEC} stellar evolution tables \citep{Bressan2012,Tang2014} to determine individual stellar lifetimes and properties which are used to set their LW band, far ultraviolet (FUV) band, and ionizing radiation luminosities (see below).

\paragraph{Stellar Feedback and Stellar Yields} We track the feedback and yields of 15 metal species for each star individually. Stars between 8 M$_{\odot} < M_{*} < 25 M_{\odot}$ explode as core collapse SNe at the end of their life, injecting their mass and 10$^{51}$~erg of thermal energy into a spherical region with radius of 5.4~pc, or 3 times the maximum resolution. Stars above this mass are assumed to direct collapse with no mass or energy injection. For all stars above 8~M$_{\odot}$ we follow their stellar winds assuming continuous mass loss over their lifetimes, their LW and FUV radiation as optically thin radiation which contributes to H$_2$ dissociation and photoelectric heating respectively, and their H~{\sc i} and He~{\sc i} ionizing radiation using an adaptive ray-tracing radiative transfer method \citep{WiseAbel2011}. We interpolate over the OSTAR2002 \citep{Lanz2003} grid to set the luminosities of each of these stars. Low mass stars, $M_{*} < 8~M_{\odot}$, do not produce feedback during their main sequence lifetimes, but end their lives injecting a short-lived, low velocity (10 km~s$^{-1}$) AGB wind \citep{Goldman2017}. Stars with $3 < M_{*} < 8$ are tracked after their death as possible Type Ia SN progenitors, using a delay time distribution model to assign when (if at all) they will explode as a Type Ia, injecting 10$^{51}$~erg of thermal energy with yields. Stellar yields are computed using the NuGrid stellar yield database \citep{Ritter2018} for all stars with $M_{*} < 25~M_{\odot}$, Slemer et. al. in prep for the stellar winds of stars with $M_{*} > 25~M_{\odot}$, and \cite{Thielemann1986} for Type Ia SNe.

\paragraph{Initial Conditions} Our dwarf galaxy is initialized to approximate, but not reproduce, the $z = 0$ properties of an ultrafaint dwarf galaxy (UFD) as informed by the observed properties of Leo P \citep[see ][]{Giovanelli2013,McQuinn2015a,McQuinn2015}. We initialize a M$_{\rm gas} = 1.8 \times 10^{6}$~M$_{\odot}$ disk as an exponential profile with a metal mass fraction of $Z = 4.3 \times 10^{-4}$ centered on a static \cite{Burkert1995} dark matter potential with M$_{\rm vir} = 2.5 \times 10^{9}$~M$_{\odot}$. The gas scale radius and scale height are set to 250~pc and 100~pc respectively, with a maximum radial extent of 600~pc. Both the gas temperatures and velocities are set iteratively to enforce initial hydrostatic equilibrium. The galaxy contains no initial background stellar population, limiting the number of Typa Ia SNe in our model. We include initial SN driving to limit the transient burst of star formation from the initial cooling and collapse of the galaxy. These SNe are randomly distributed with the same radial and vertical scale heights as the gas disk at a rate of 0.4~Myr$^{-1}$; this corresponds to the current SFR of Leo P, $\sim 4\times 10^{-4}$~M$_{\odot}$~yr$^{-1}$ \citep{McQuinn2015a}, assuming 1 SNe per 100 M$_{\odot}$ of star formation. The metal yields from these SNe are set to the mean ISM abundances, and thus do not contribute to the chemical evolution of the galaxy. Although the total metallicity field is initially non-zero, we only track the self-consistently produced metal enrichment for each individual metal species in our simulation, setting their initial abundances to zero. Our analysis is based on the evolution of this galaxy during the first 500~Myr after the formation of the first star particle.

\section{Results}
For context, in Paper I we focused on the global properties of the evolution of this dwarf galaxy, including an analysis of the galaxy's gas mass and star formation evolution, the ISM properties in terms of mass fractions, volume fractions, and phase diagrams, the interstellar radiation field in each tracked radiation band, the gas outflow rates and galactic wind velocities, and the retention / ejection of metals from the galaxy. This galaxy has an average star formation rate of $1.2~\times 10^{-4}$ M$_{\odot}$ yr$^{-1}$ and is consistent with observed low mass dwarfs galaxies in the Kenicutt-Schmidt relation.  The galaxy exhibits strong, feedback-driven outflows \citep{Emerick2018b} that eject a significant amount of gas and metals from the galaxy. The mass loading factor at 0.25~R$_{\rm vir}$ was found to be $\eta \sim 50$, where $\eta$ is defined as the mass outflow rate in through a 0.1~$R_{\rm vir}$ annulus centered at 0.25~R$_{\rm vir}$ divided by the 100~Myr averaged SFR. These winds eject 96\% of the metals produced in the galaxy, with 50\% of all metals leaving the virial radius by the end of the simulation time.

In this work we focus in detail on how each of the 15 individual metal species evolve in this galaxy. We address differences between the ejection fractions of each metal in Section~\ref{sec:ejection} and analyze for the first time the mass-fraction PDFs of the metals retained by the ISM in Section~\ref{sec:mixing}.

\subsection{Preferential Ejection of Metals from the ISM}
\label{sec:ejection}

\begin{figure*}
\centering
\includegraphics[width=0.99\linewidth]{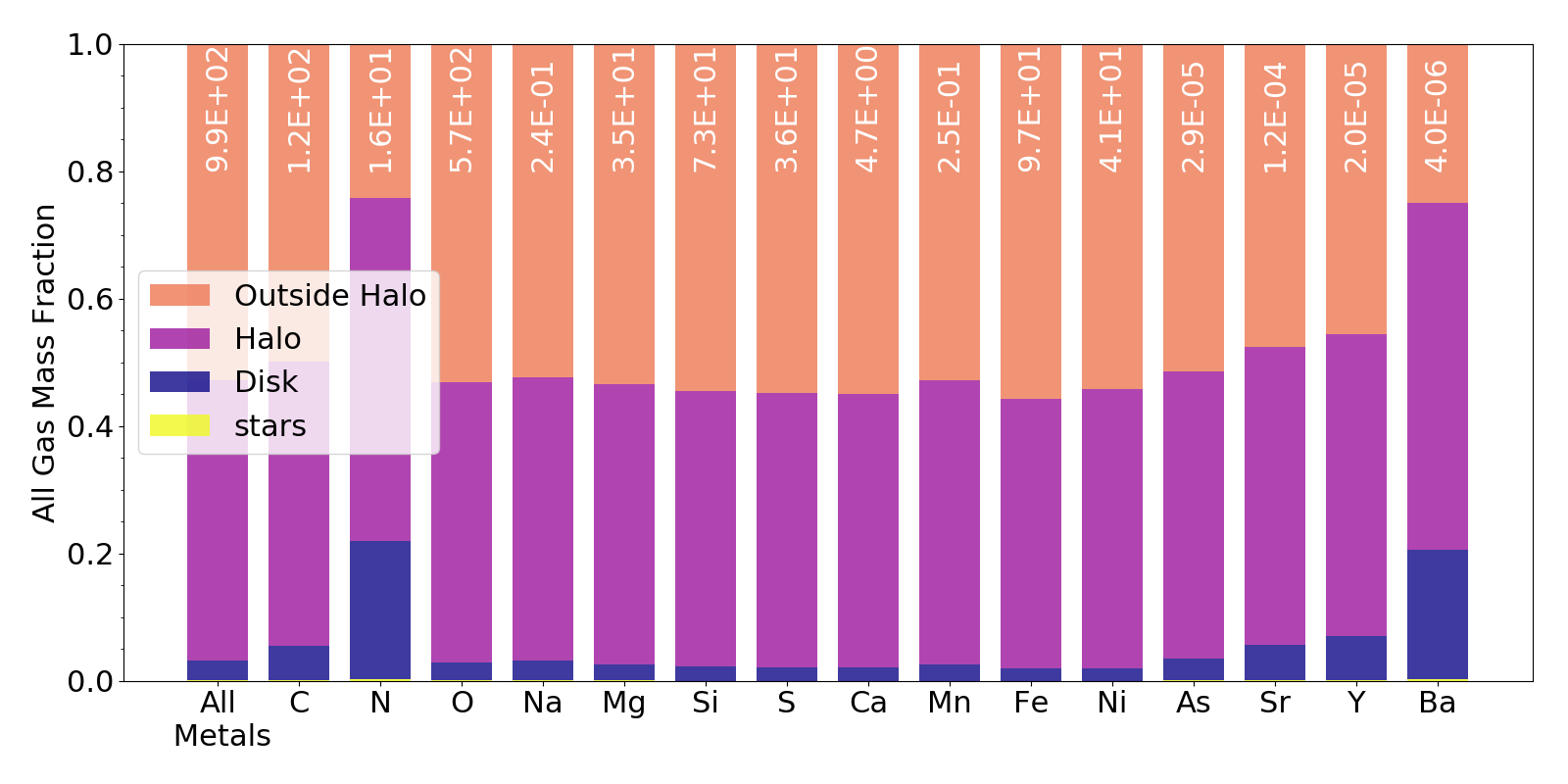}\\
\includegraphics[width=0.99\linewidth]{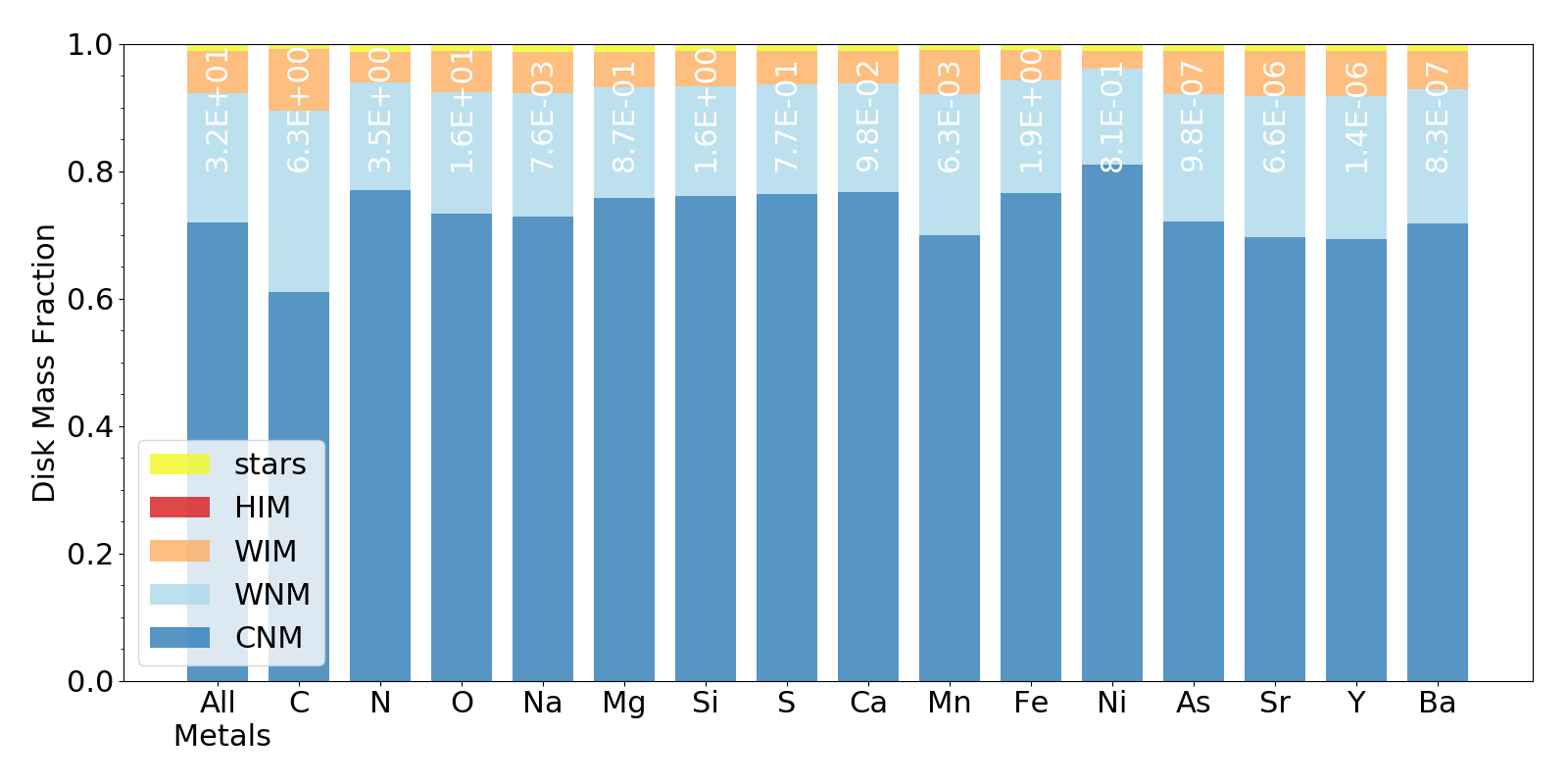}
\caption{Top: The mass fraction of each metal species in the \textit{full simulation box} contained in the halo, gas in the galaxy disk, and stars. Bottom: The mass fraction of species within \textit{the disk} alone in each phase of the ISM (bottom). The leftmost bar in each plot shows the sum of all metals.}
\label{fig:species_fractions}
\end{figure*}

%mm [this all seems to have been deleted in one version, but doesn't look like it should have been.  Down to ">>>>"] <<<<<<< HEAD
The top panel of Figure~\ref{fig:species_fractions} gives the mass fraction of each element at the end of the simulation (500~Myr) in each of four reservoirs: locked in stars (yellow), in the ISM (blue), outside the galaxy but within the virial radius (purple), and outside the virial radius (salmon), including gas that has left the domain. We subdivide the ISM by phase in the bottom panel, giving the mass fractions of each element in the cold neutral medium (CNM, $f_{\rm H_2} < 0.5$,  T$< 100$~K), warm neutral medium (WNM, $10^2~\rm{K}\le \rm{T} < 10^4~\rm{K}$), warm ionized medium (WIM, $10^4~\rm{K}\le \rm{T} < 10^{5.5}~\rm{K}$), the hot ionized medium (HIM T$\ge 10^{5.5}$~K), and locked in stars (yellow).\footnote{The $f_{\rm H_2}$ restriction on the CNM is not relevant in our simulations, as there are no cells with $f_{\rm H_2} > 0.5$. $f_{\rm H_2}$ remains below about 0.35 for all cells (see Paper I).}
%mm Again, with the exception of  H and He, we [H & He aren't metals]
    We
only consider metals produced self-consistently through our star formation and stellar feedback methods; the initial mass of each metal species is zero.

We find two major results.
%mm [reversed order of results to follow figure.  The opposite choice could also be made, by reversing the order of the figure...]
    First, 
only a small fraction of produced metals are retained within the dwarf galaxy, in agreement with observations of nearby dwarf galaxies \citep[see][]{Kirby2011-metals, McQuinn2015}; however, the retention factor varies 
%mm 
    among elements. The top panel shows a qualitative disagreement between the retention fractions of N and Ba (about 20\% for each) as compared to the rest of the metals ($\sim$ 4 -- 5\%). This suggests that individual metals \textit{do not} share the same dynamical evolution. 
%mm Clearly
       Thus, 
metal enrichment in galaxies is a phenomenon that cannot be fully captured using a single, global metallicity field. In this particular case, assuming that all metals behave the same would underestimate the N and Ba enrichment by a factor of up to five, at least for low mass dwarf galaxies.

    Second,
nearly all of the metals 
%mm
    retained in the disk
reside within neutral gas, mostly in the CNM; only a few percent
reside in the hot phases. The exact fraction varies with each species, most notably for carbon; these fluctuations are at most $\sim 10$\%. This is not surprising, as the cold phases represent the majority of the mass in the ISM, but, as shown in Section~\ref{sec:statistics}, even though the cold phases harbor most of the metals, the hot phases have significantly higher metal mass fractions.

\begin{figure*}
\centering
\includegraphics[width=0.95\linewidth]{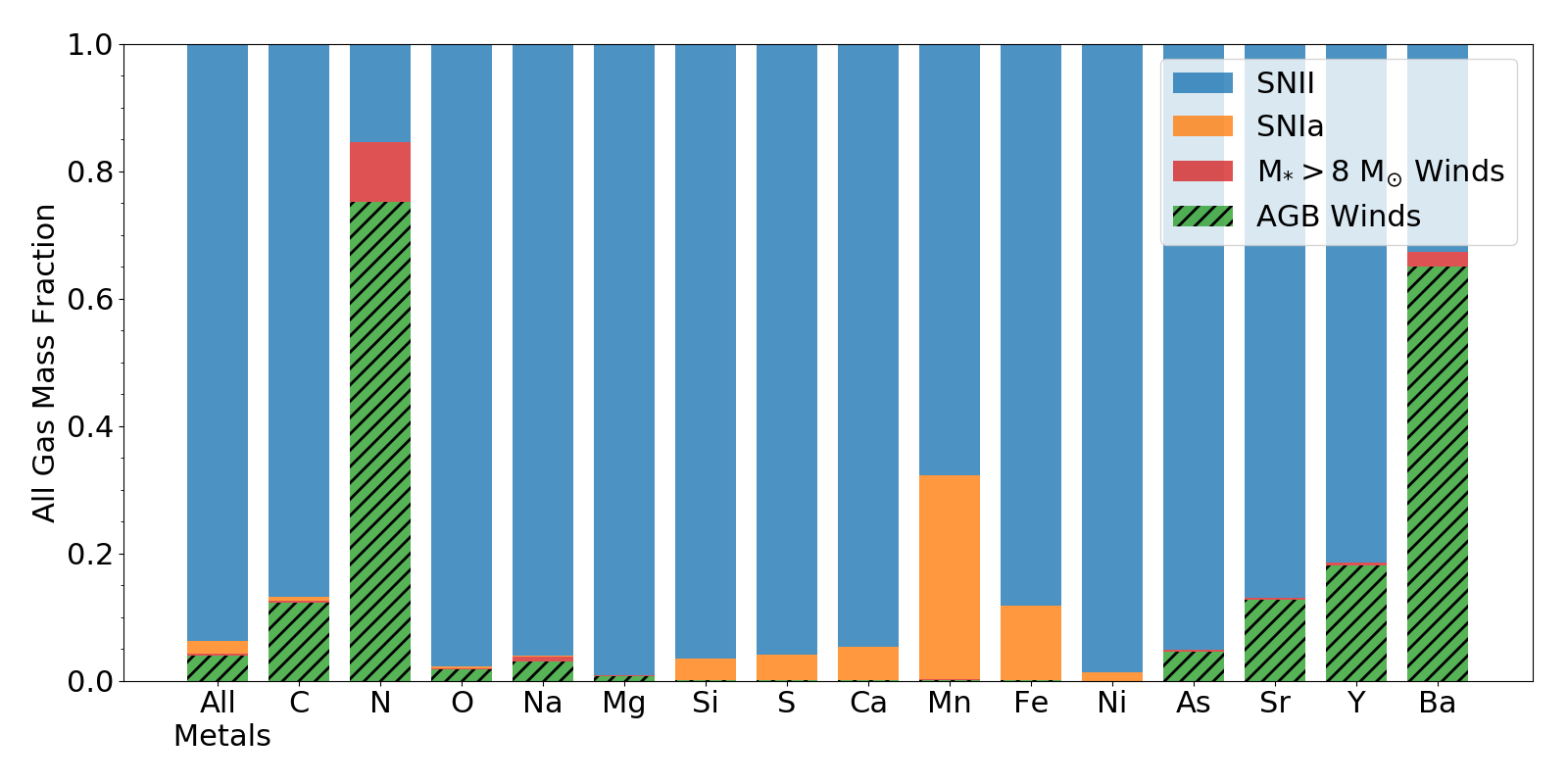}\\
\caption{The fraction of total mass in each metal species produced by each of the four possible nucleosynthetic channels in our model. These channels differ in both when metals are ejected, as determined by stellar evolution, and the phase of the ISM into which they are ejected.
We note that the minimal contribution from Type Ia SNe for the iron-peak elements is because  no older stellar population was initialized, so only 16 of them have exploded by the end of our 500~Myr simulation.}

\label{fig:species_sources}
\end{figure*}

The only physics that separates the dynamical evolution of these elements in our simulations is the individual sources of enrichment: AGB winds (stars less than 8~M$_{\odot}$), stellar winds (stars above 8~M$_{\odot}$), core collapse SNe, and Type Ia SNe. These channels differ in: 1) how long after a given star formation event they occur, 2), by consequence, the typical ISM properties in which they occur, and 3) how much energy is associated with each event, which determines ejecta temperature and velocity. To understand where each of the elements in Figure~\ref{fig:species_fractions} originated, we show the mass fraction of metals produced through each channel in Figure~\ref{fig:species_sources}. Core collapse SNe are responsible for over 90\% of the total metal enrichment in our galaxy, but this is clearly not true for all elements. In particular, a majority of N (74\%) and Ba (65\%) are ejected by AGB winds; both are retained at a higher rate than the rest of the metals. That this behavior exists for N and Ba in our simulations is dependent upon the metallicity of our galaxy and choice of stellar yield tables, but we can generalize this result to say that, for any assumed set of yields, low mass galaxies should more easily retain \textit{any} elements synthesized predominately in AGB winds, as compared to elements synthesized through SNe.\footnote{One might expect that much of the N and Ba that is ejected by galactic winds is comprised mostly of the $\sim$30\% of each species produced through the stellar winds of more massive stars and SNe. However, this cannot be proven as we lack Lagrangian information about individual gas elements.}
% AE: Actually if there are significant isotope variations between nucleosynthetic sources (i.e. N produced in AGB winds is a different isotope than the trace N produced in SNe) then that could be a very cool thing to distinguish if we had good gas-phase abundance measures of various isotopes... but maybe not if they aren't very stable....
We discuss how this result may extend towards other metal yields that are not well sampled on our relatively short (500 Myr) simulation timescales in Section~\ref{sec:discussion:metal yields}.

\begin{figure}
\centering
\includegraphics[width=0.95\linewidth]{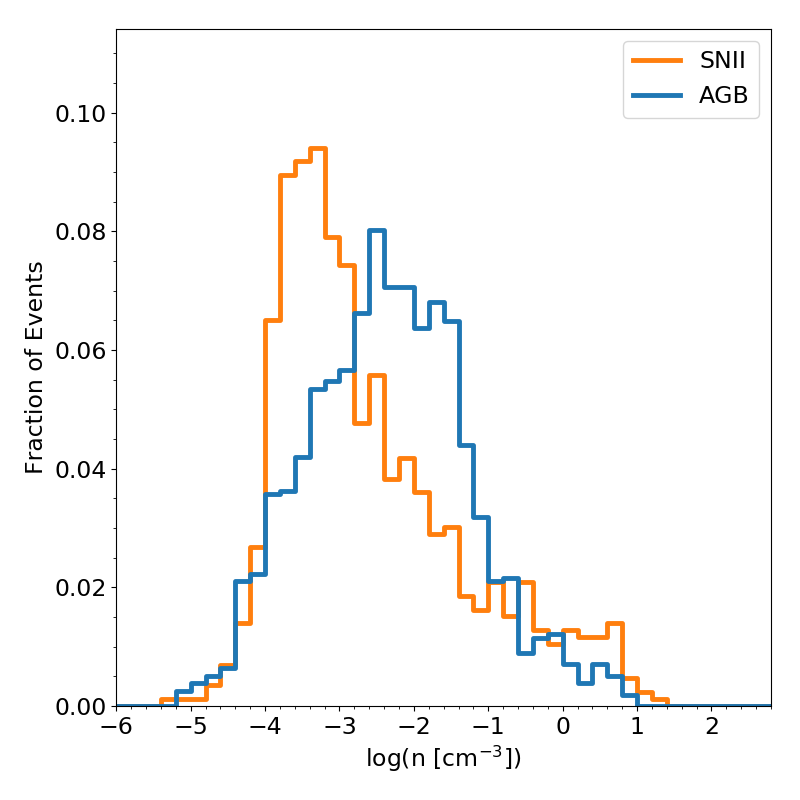}\\
\includegraphics[width=0.95\linewidth]{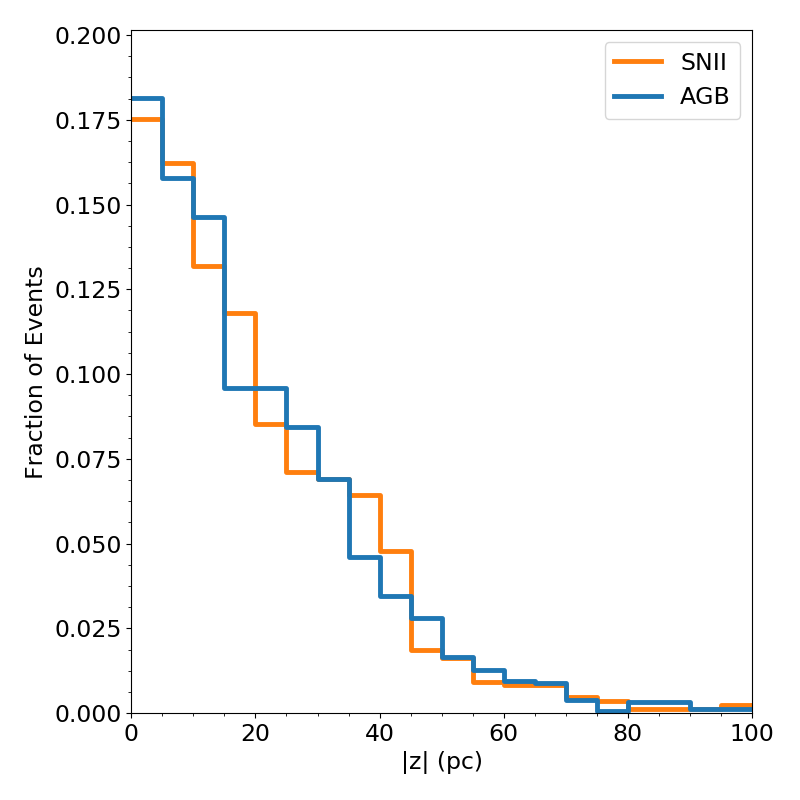}
\caption{The volume-averaged gas number densities within 20~pc of a given event (top) and vertical position above/below the disk (bottom) within 1~Myr before the event.}
\label{fig:spatial distribution}
\end{figure}

AGB winds have low energy and velocities (10~km~s$^{-1}$) as compared to the energy and typical expansion velocities of SNe ($\sim$1000~km~s$^{-1}$). In addition, their longer timescales relative to massive stars means that AGB stars are typically removed from their birth regions and randomly distributed through the galactic disk. The changes in typical ISM density and height of these events contributes to these variations. We show histograms of the average number density within 20~pc of any given enrichment source (top) and height above/below the disk (bottom) within 1~Myr of each event (as limited by our output cadence) in Figure~\ref{fig:spatial distribution}. As shown, SNe peak at very low densities, indicating that most explode in superbubbles, regions carved out by previous SNe. AGB stars predominately release their metals close to the average ISM density. The scale height distributions for both events show no significant differences. We do not expect these differences to be the dominant effect in driving the differential evolution of elements ejected by AGB winds vs. SNe, compared to the energetics, but can certainly play a significant role in determining the mixing behavior of individual enrichment events. The changes to mixing behavior as a function of ISM properties will be investigated in more detail in a future work.

\subsection{Mixing and Distribution of Metals in the ISM}
\label{sec:mixing}
We characterize the metal distributions (in Sections~\ref{sec:log-normal} and \ref{sec:phase-pdfs}) and evolution (in Section~\ref{sec:statistics}) in our galaxy to build towards a  a complete model for how metals mix and evolve in the complex, multi-phase ISM of real galaxies.

\subsubsection{A Functional Form for Metal PDFs}
\label{sec:log-normal}

The log-normal distribution is found often in nature, generally describing multiplicative processes with non-negative values that grow with time. In astrophysics, for example, the log-normal distribution can be used to describe the time evolution of the star formation rate density \citep[see ][]{Gladders2013,Abramson2016,Diemer2017}. In addition, as expected from analytic theory \citep{Vazquez-Semadeni1994}, isothermal turbulence gives rise to log-normal density probability distribution functions \citep[PDFs;][]{Padoan1997, Passot1998, Ostriker1999,PadoanNordlund2002,KrumholzMcKee2005,Federrath2008}. Although these PDFs are only log-normal in simulations containing a more realistic, multi-phase ISM \citep{Scalo1998} if the disk is very stable \citep{WadaNorman2007}, individual phases within the ISM do exhibit some log-normality \citep{Tasker2009, Tasker2011,Joung2009,PriceFederrathBrunt2011, HopkinsQuataertMurray2012}. It has been shown that the 3D density PDF and the column density PDF, in both simulations and observations, have a characteristic shape \citep{Vazquez-Semadeni1994,Burkhart2009, FederrathKlessen2013, Collins2012, Myers2015, Burkhart2017, Chen2018}. This includes a log-normal component, generated by multiplicative processes---in this case shocks and the turbulent cascade in the ISM---and a power-law component at high densities arising from the additive combination of individual, self-gravitating cloud structures .

The physics that drives the density PDF is directly related to the process of metal mixing and diffusion. However, there is no a priori reason why gas density and metallicity PDFs should have similar functional forms. We demonstrate here for the first time that the mass fraction PDFs for each metal species in our simulation can indeed be well fit using a piecewise log-normal + power-law PDF. We use a simple conceptual model in Section~\ref{sec:interpretation} to motivate the emergence of this distribution.

We follow \citet{Collins2012}, \citet{Burkhart2017}, and \citet{Chen2018} in constructing our piecewise PDF. We define the distribution of metals in the ISM as a function of the fraction of mass contained at a given metallicity, $Z$, or metal mass fraction, $Z_i$, where $i$ denotes an individual element. This PDF is given as
\begin{align*}
  p(Z) =
  \begin{cases}
    \frac{N}{\sigma Z \sqrt{2\pi}} \rm{exp}\left[-\frac{(\rm{ln}(Z) - \mu)^2}{2\sigma^2}\right],
    & Z < Z_{\rm{t}} \\
    % \multicolumn{1}{@{}c@{\quad}}{1} % variant with \multicolumn
    N p_o Z^{-\alpha},
    & Z > Z_{\rm{t}}
\end{cases}
\end{align*}
where $N$ is a normalization constant, $p_o$ ensures continuity between the two components, $\mu$ and $\sigma$ are the log-mean and width of the log-normal component, $\alpha$ is the power-law slope,  and Z$_{\rm t}$ is the metal fraction at the transition between the log-normal and power-law components. When fitting this PDF, we only enforce continuity and $\int_0^{\infty} p(Z) dZ = 1$, leaving $\alpha$, $\mu$, $\sigma$, and $Z_{\rm t}$ as free parameters. These conditions set $N$ and $p_o$ to
\begin{equation}
N = \frac{1}{2} \left[ 1 + \rm{erf}\left( \frac{ln\left(Z_{\rm t}\right) - \mu}{\sigma\sqrt{2}}\right)\right] + \frac{p_o}{\alpha-1}Z_{\rm t}^{-\alpha + 1}
\end{equation}
and
\begin{equation}
p_o = \frac{1}{\sigma Z_{\rm t} \sqrt{2 \pi}} {\rm exp}\left[-\frac{({\rm ln}(Z_{\rm t}) - \mu)^2}{2 \sigma^2} + \alpha {\rm ln}(Z_{\rm t}) \right]
\end{equation}.

We compute the numerical metal mass fraction PDFs for each metal species in our simulation using a fixed bin 
%mm with 
         width
of 0.05 dex. We fit $p(Z)$ to each of these using a Levenberg-Marquardt algorithm as implemented in \textsc{SciPy} \citep{SciPy}, stepping through possible values for $Z_{\rm t}$, set to the centers of each of these bins. The best of these fits is then compared to best fits using only a log-normal component or only a power-law component, and the best of these three is accepted. The log-normal + power-law PDF produces the best fit in nearly all cases.

\begin{figure*}
\centering
\includegraphics[width=0.95\linewidth]{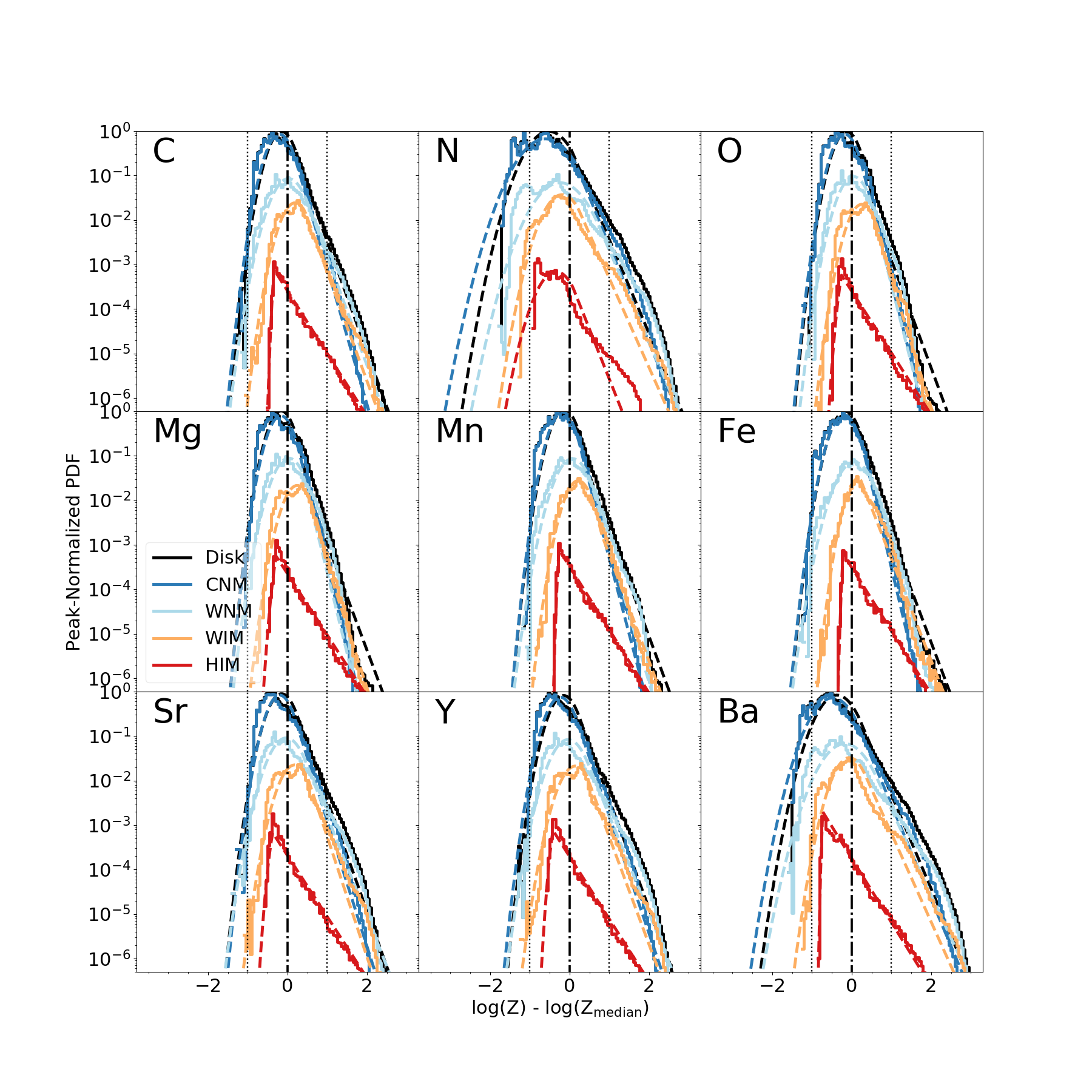}
\caption{The numerical PDFs (solid histograms) and the associated log-normal + power-law fits (dashed lines) for a subset of the elements tracked in our simulation in each of the four gas phases defined in Section~\ref{sec:ejection}: CNM (dark blue), WNM (light blue), WIM (light orange), HIM (red), and all the gas in the ISM (black). For clarity, each distribution is normalized to the mode of the full-disk PDF (black) and is centered on the median value of the full-disk PDF. We note the vertical axis normalization is such that integrating over the shown PDF gives the mass fraction of that phase in the disk. Since the CNM dominates the mass fraction of our galaxy, the black curve is often obscured at low metal fractions.}
\label{fig:log-normal}
\end{figure*}

We show in Figure~\ref{fig:log-normal} the numerical PDFs (solid histograms) and log-normal + power-law fits (dashed lines) across individual gas phases. These PDFs have been computed at an arbitrary single point in time in the middle of the simulation run.
% 270 Myr
For clarity, we only show a subset of the 15 elements we follow. As shown, there are clear differences in the PDFs across elements of different nucleosynthetic sources (AGB vs. SNe) and between each phase. However, each of these distributions are characterized by a power-law tail towards high metal fractions and a turnover of varying width at low metal fractions. We discuss the differences among each phase in more detail in the next section, but note here that the significance of each of these components varies notably across phases. In many of these cases, the piecewise log-normal + power-law distribution fits the numerical PDF quite well. However, there are often deviations, particularly at low metal fractions, from pure log-normal behavior. This manifests as either a very broad, flat PDF at low metal fraction (see N and Ba, for example), or large peaks not well described by $p(Z)$ at low metal fraction. In these situations, it is unclear what, if any, portion should be considered as a log-normal, or if there are multiple components within this region.

We argue here that the log-normal + power-law PDF can be a powerful tool for modeling the metal fraction  PDFs of individual elements in galaxy models. The fits are not uniformly perfect but some deviation from a simple analytic model is expected in a complex, multi-phase ISM. In addition, some of this deviation, particularly in the WNM and WIM, could be caused by grouping together qualitatively distinct gas in a single phase; this would tend to broaden the distributions. Most importantly, however, the PDFs of the CNM, are indeed well fit by the adopted $p(Z)$. As this is the source of star-forming gas in the ISM, the log-normal + power-law PDF appears useful to account for intrinsic scatter in stellar abundances in galaxy evolution models.

\subsubsection{PDF Variation Across Gas Phase}
\label{sec:phase-pdfs}

The various phases represented in Figure~\ref{fig:log-normal} involve variations in density and temperature of more than six orders of magnitude. The evolution of each phase is qualitatively different, and the metal mixing behavior of each should vary. Mixing timescales over a given length scale should be related to the local sound speed; hot gas, with higher sound speeds, should mix more rapidly than the dense, disconnected clumps of cold gas in the ISM. In addition, one would expect a metallicity gradient with gas temperature as enrichment occurs first in the hot phases, cooling and enriching denser gas over time. We examine the PDF variations among elements in each phase, as shown in Figure~\ref{fig:log-normal}, in more detail here.

Unsurprisingly the diffuse HIM is the most metal rich phase, as it is comprised predominately of metal enriched SN ejecta. Clearly this leads to long power-law tails towards high metal fractions for each species, with a very narrow, poorly defined peak at low metal fractions. Generally, in colder gas the metal PDFs become less enriched with broader, low-metal-fraction components and steeper, power-law tails. Although the extended power-law tail of the HIM leads to a large range of metal fractions, the HIM represents very little mass, and this tail represents recent, un-mixed enrichment. The comparatively narrow width in the log-normal regimes of the CNM is perhaps surprising. Although this gas is at lower metal fraction than the WNM and WIM, it would appear that it is more well mixed. This runs counter to the idea that mixing times should be long in colder gas, unless (as we argue) mixing first proceeds rapidly in the hot phases before mixing in with cold, disconnected structures across the galaxy.

Although these trends across phases hold for all elements, there are qualitative differences between elements, particularly between those ejected predominantly by AGB winds (e.g. Ba and N) and those ejected by SNe (e.g. O and Mg). In all phases, except the HIM, the AGB-wind elements have broader distributions that are less well described by our adopted $p(Z)$ than the metals dominated by SN enrichment. The power-law component of N and Ba is generally shallower than all other metals across each phase (except the HIM), particularly in the CNM. Ba and N do not show significant differences among the rest of the metals in the HIM, though this is likely because the Ba and N present in this phase is dominated by the Ba and N included in SNe yields. Again these differences between yield sources could be driven both by differences in their enrichment timescales and therefore differences in the typical ISM environment each event encounters, and the differences in energetics between AGB winds and SNe. AGB wind elements enrich the WNM and WIM directly, rather than the HIM. This leads to longer mixing timescales and broader PDFs for these elements.

\subsubsection{The Time Evolution of Metal PDFs}
\label{sec:statistics}
\begin{figure*}
\centering
\includegraphics[width=0.95\linewidth]{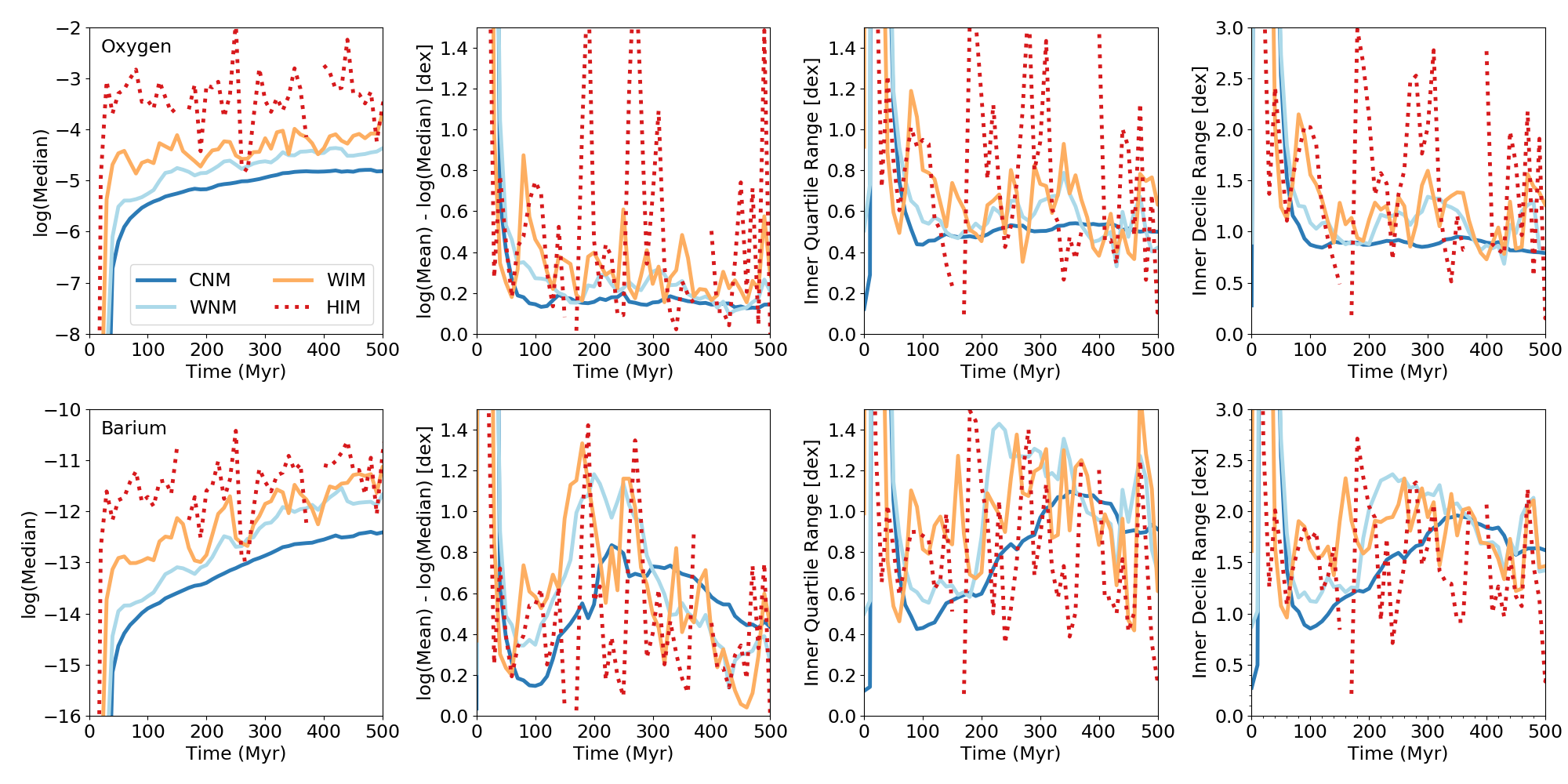}
\caption{Time evolution of three different statistics for the full distributions of O (top) and Ba (bottom) in each phase of our simulation. The panels show log$_{10}$ of the median (left), the difference, in dex, between the mean and the median (middle), and the 90$^{\rm th}$ decile and 10$^{\rm th}$ decile difference in dex (right).}
\label{fig:phase-statistics}
\end{figure*}

We focus on the time evolution of the full numerical PDFs in this section. Our results here do not depend upon the choice of functional form for $p(Z)$. Figure~\ref{fig:phase-statistics} shows the evolution of four different statistics for the O (top) and Ba (bottom) PDFs. These two elements are treated as representative elements for SN and AGB wind production respectively. The difference between the mean and median masses of the PDF (second column) is a measure of the skew of the PDF. Positive values indicate that metals are preferentially sequestered in metal-rich gas, and are less well mixed throughout the given phase. The skew is always positive in these distributions.

For O, the median mass fraction is ordered by phase temperature. The HIM is significantly more enriched than the cooler phases by anywhere from 0.1 dex to 4 dex, fluctuating by $\pm 1$ dex over the simulation time. The frequent large skew in the HIM (see second panel) and spread in the HIM (right two panels), coupled with its continual fluctuation, suggests that the HIM is not in equilibrium.
%This phase is poorly mixed because it is short lived, either cooling into different phases of the ISM or being ejected from the galaxy, and is continually enriched by new SN events.
Each cooler phase in O is progressively less enriched (lower median), with smaller skew and spread. The offset between phases and increasingly well mixed gas from hot to cold indicates that metal enrichment in the ISM of galaxies proceeds first through mixing on large scales in a hot phase, before progressively cooling and/or mixing through multiple phases until enriching star-forming gas. This 
%mm is the only explanation for [that's perhaps too strong a claim]
    can explain
how the cold gas can rapidly homogenize over the whole galaxy within $\sim$50 Myr, roughly when the cold phase exhibits a nearly constant spread (right panels). Individual enrichment events 
%mm will 
have significantly higher metallicities than the ambient ISM in any phase, and 
%mm will 
drive an increase in the difference between the mean and median of the PDF. These can be seen as the obvious spikes in the HIM and WIM. For O, the lack of these spikes in the two cold phases suggests again that enrichment does not occur directly in these phases, but proceeds more gradually through the warmer phases first.

% In each phase, enrichment events will tend to drive larger differences between mean and median mass fraction,

Although these trends are generally true for Ba, its evolution is much more complicated.
%In addition, the HIM is much less offset from the rest of the phases than is seen in O, with much less variability.
Unlike O, the spread and skew in Ba for the CNM, WNM, and WIM \textit{increase} during the evolution, reaching differences in 90$^{\rm th}$ and 10$^{\rm th}$ percentiles of nearly 2 dex in the CNM and over 2 dex in the WNM and WIM. The HIM is seemingly unaffected by this trend, and simply fluctuates throughout the simulation. Given their lower energies, AGB winds more directly enrich the WNM and WIM, not the HIM as in SNe. The consequence of this is clear in Figure~\ref{fig:log-normal} by the wider PDFs and longer power-law tails in Ba in these phases. These tails represent the most recently enriched gas, 
%mm and 
        which
is clearly much more locally confined than O.

%{\bf (move to discussion?)}:
The positive skew in all of the PDFs presented here implies that most of the mass of the galaxy has a metal fraction below what one would normally adopt as the average metallicity (i.e.\ the ratio between the total mass of metals and the total mass). Assuming star-forming gas follows the same properties as the cold gas, this distinction is small ($\sim$ 0.2 dex), 
%mm but
   though 
significant, for elements released during SNe explosions, but can be very significant, up to $\sim$ 0.8 dex, for elements released in AGB winds. Chemical evolution models, especially one-zone models, follow the mean metal fraction, rather than the median. 
%mm These results indicate that these 
            Our results indicate that such
models are biased to overestimate gas and stellar elemental abundances.
% However star formation may not be unbiased, and may preferentially come from gas sitting on the higher metal fraction tails.

To summarize, Figure~\ref{fig:phase-statistics} demonstrates: 1) there are qualitative differences in how SN-injected elements (e.g. O, Mg) and AGB-wind-injected elements (e.g. N, Ba) are distributed through the ISM, with the latter having a broader range of variation and being less well mixed in all phases except the HIM; 2) hotter phases are more metal enriched, both because the cooler phases make up most of the initially unenriched mass of the ISM and because the hotter phases are more directly populated by recent enrichment events; 3) the cooler, denser phases, particularly for SN injected elements, are more well mixed than the hot phases of the ISM; 4) the PDFs of metal mass fraction are best fit by a log-normal + power law function; and therefore 5) the median metallicity available for star formation 
%mm rates 
lies below the mean galactic value. In the case of SN injected elements, enrichment proceeds quickly through the HIM over the entire galaxy. For AGB injected elements, enrichment proceeds through the WIM and WNM, leading to longer mixing timescales and larger metal fraction variations in the ISM.

\section{Discussion}
\label{sec:discussion}
We begin with a simple toy model that motivates the power-law tail at high metal fractions of the metal fraction PDFs and a subsequent turnover at low metal fractions in Section~\ref{sec:interpretation}. This work is placed in context with previous papers focusing on metal mixing in the ISM in Section~\ref{sec:context}. We discuss the relevance of additional AGB yields not directly followed in this study in Section~\ref{sec:discussion:metal yields}. Finally, in Section~\ref{sec:stellar abundances} we discuss how these results relate to stellar abundances, make generalizations to more massive galaxies in Section~\ref{sec:massive galaxies}, and discuss possible impacts of these results on chemical enrichment from more exotic nucleosynthetic sources in Section~\ref{sec:exotic enrichment}.

\subsection{Physical Interpretation of the PDF}
\label{sec:interpretation}
%
% KVJ wanted clarification on this argument. I believe my added text 
% does this sufficiently.
Take the simple case of an initially primordial, uniform, isothermal medium of mass $M_o$ with initial metallicity $Z_{\rm o}$, containing a single, un-mixed enrichment event of mass $M_{\rm ej}$ whose size is small compared to the system and mass $M_{\rm ej} / M_o \ll 1$. Thus the background medium represents a virtually inexhaustible (but finite) source of un-enriched gas. In this case, $p(t,Z)$ initially takes the form of a double-delta function
\begin{equation}
p(t_{\rm o},Z) = \delta(Z_{\rm o}) + \frac{M_{\rm ej}}{M_o}\delta(Z - Z_{\rm ej}).
\end{equation}
%mm [edited the rest of this paragraph, and the next though I don't think there are any substantive changes unmarked.]
If the enriched material mixes continually with the un-enriched gas at a constant rate, after some time  $\tau_{\rm mix}$ this gas will have mixed with an equal amount of primordial gas. At this time, $p(\tau_{\rm mix},Z) = \delta(Z_{\rm o}) + (2M_{\rm ej}/M_o) \delta(Z - (1/2)Z_{\rm ej})$. Then, $p(2\tau_{\rm mix},Z) = \delta(Z_{\rm o}) + (4M_{\rm ej}/M_o)\delta(Z - (1/4) Z_{\rm ej})$, and so on for other multiples of $\tau_{\rm mix}$. This represents an inverse relationship between the mass of enriched gas and the metallicity of the enriched gas. The superposition of the distributions of a single event evolving in time will appear as a power-law distribution with slope $\alpha = 1$.

Thus, if the gas is continually enriched by identical, well-separated enrichment events of mass $M_{\rm ej}$ and metallicity $Z_{\rm ej}$, the instantaneous distribution $p(t,Z)$
%mm , or $p(Z)$, 
of these many events will be $\delta(Z_{\rm o})$ plus a power law in $Z$ with slope $\alpha = 1$, truncated at some minimim $Z$ (related to the time since enrichment began) and some maximum, $Z_{\rm ej}$.
% and thus a power-law evolution in time of the high metallicity portion of $p(Z)$ with slope $\alpha = 1$. 
The slope of the power-law, then, is determined by the rate of injection versus mixing with the ambient medium. A power-law index $\alpha < 1$ can occur when injection occurs more rapidly than the newly enriched gas can mix with the ambient medium. Steeper power-laws, $\alpha > 1$, develop when mixing occurs more rapidly than 
%mm enrichment. 
         injection.
If the ambient, primordial gas were truly infinite, the power-law would never completely encompass the delta function at $Z = Z_o$.

In reality, however, $M_o$ is not an inexhaustible reservoir of primordial gas. Eventually the entire ambient medium will become enriched to some $Z > Z_o$, and $p(Z)$ will consist entirely of a truncated power-law. As enrichment proceeds, gas near the low metallicity truncation of $p(Z)$, which still comprises much of the mass of the system, enriches towards higher metallicities in the power-law tail. This will produce a turnover at the low $Z$ limit of $p(Z)$. The low-turnover limit is produced by diffusion from many different sources, and is thus a multiplicative process, which, as we noted above, tends to produce log-normal distributions. 
%mm [additive vs multiplicative]
   %{\bf [According to Section~\ref{sec:log-normal}, MULTIPLICATIVE processes produce log-normals. Additive processes produce power-laws.  However, see Mouri (2013) for a possible counterexample...]}
The physical interpretation of these two components is that the power-law tail represents newly enriched and poorly mixed gas that is above the average gas metallicity and is undergoing dilution, while the log-normal component represents the ambient medium that lies below the average gas metallicity and is undergoing enrichment.

This toy model provides a physical intuition for the general trends in the PDFs presented here across ISM phases. 
%mm The HIM is comprised predominantly of gas from individual enrichment events that, due to its low density, easily create large power-law tails beginning at very high metal fractions. 
     Individual enrichment events start at very high metal fractions in the low-density HIM, 
     so they easily create long power-law tails.
Whatever ambient component of the HIM that exists is well mixed, leading to a narrow and sub-dominant log-normal component at low metallicities. Cold, dense gas, which is almost never directly impacted by these individual enrichment events, is enriched almost entirely by diffusion, and thus has an almost completely log-normal PDF with very little, if any, power-law tail.

%
%
%
%Although this model is useful to explain these very general features, it is certainly incomplete in practice. Enrichment proceeds irregularly through the mulit-phase ISM with a variety of ejection masses and metallicities. Phases are not independent, and the hot phase cools and enriches cooler phases in the ISM, distorting $p(Z)$ within each phase. Cold gas is heated up and destroyed through feedback, changing the content of the low-metallicity end of the warm and hot phases. Finally, gas is removed from the galaxy through feedback driven winds that preferentially remove metal enriched gas. Each of these processes has a non-trivial effect on the evolution of the PDF that will cause deviations from the simple toy model outlined here. Determining the exact physical processes that drive and set the properties of each PDF is the subject of ongoing work.

%
% A.E. paragraph may need work
%

Mixing timescales are likely proportional to the eddy turnover times at the injection scale \citep{PanScannapieco2010, Colbrook2017} and the properties of turbulence in the ISM \citep{YangKrumholz2012,Sarmento2017,Sarmento2018}. Mixing within a phase is likely dependent upon the phase's sound speed and velocity dispersion. This would imply rapid mixing timescales in the WIM and HIM, with typical velocity dispersions of $\sim$~30 km~s$^{-1}$ and $\sim$~100~km~s$^{-1}$ respectively, and long mixing timescales in cold, dense gas ($\sim$ 1 km s$^{-1}$). Our results show, however, that in general the WIM and HIM are the least well mixed, while the CNM is the most well mixed across the galaxy. This is particularly curious as the sound crossing time of the HIM across the galaxy ($\sim 1$ kpc) is $\sim$10~Myr, compared to $\sim$~1~Gyr for the coldest gas. It must be that the HIM is far from equillibrium throughout the simulation, in part due to the continual enrichment by ongoing SNe. Newly enriched gas must mix through the HIM on galaxy scales, becoming well mixed and less enriched by the time it cools into the CNM and eventually star-forming gas. Since elements produced in AGB winds do not directly enter the HIM, mixing is less efficient driving larger variations across the galaxy. The idea of metals processing first through the hot ISM has been proposed before to explain observed metallicity trends in the outskirts of more massive galaxies \citep{Tassis2008,Werk2011}.

% From benoit: more explicit title
\subsection{
%mm Context With 
        Comparison with 
Previous Studies of Metal Mixing}
\label{sec:context}
That metal fraction distributions can be described using simple analytic log-normal + power-law PDFs, even in a complex, multi-phase ISM, has not been demonstrated prior to this work. In addition, this is the first work to demonstrate differential mixing behavior of individual metal species using 3D hydrodynamics simulations. However, there does exist a significant body of work investigating the mixing behavior of passive scalars in a variety of contexts.
%mm , as discussed below.  [ moved KrumholzTing18 to end of following paragraph.]
%We discuss our results further in the context of previous research on metal mixing in the ISM and speculate on the physical processes that drive the evolution of these PDFs across phases.

Previous work on the evolution of metals in the ISM varies from studies concerning the advection of passive scalars in idealized turbulent boxes \citep[e.g.][]{Pope1991, PanScannapieco2010, PanScannapiecoScalo2012, PanScannapiecoScalo2013, YangKrumholz2012, SurPanScannapieco2014, Colbrook2017}
% cite passive scalar reviews by Warhaft 2000 in fluid mech and scalo + Elmegreen 2004?
to global galaxy models studying generalized advection and mixing of passive scalars \citep[e.g.][]{deAvillez2002,Petit2015,Goldbaum2016} to models with more detailed self-consistent metal enrichment \citep[e.g.][]{Revaz2009,Escala2018}.
\cite{KrumholzTing2018} predict differential behavior for AGB wind and core collapse SN synthesized elements as a direct consequence of the differences in size of typical planetary nebulae ($\sim$ 0.1~pc) and SN remnants ($\sim$ 100~pc).

The evolution of metallicity PDFs has been investigated previously in some of these works \citep[see ][ and references therein]{PanScannapiecoScalo2012,PanScannapiecoScalo2013}, with effort towards developing closure models to describe the evolution of the PDFs of passive scalars in turbulent media \citep[e.g.][]{EswaranPope1988,Chen1989,Pope1991}. The astrophysical context in much of this work was enrichment from the first stars, so the focus was on the low-metallicity tail of the PDF and the timescales over which gas is polluted \citep[e.g.][]{PanScannapiecoScalo2013,Sarmento2017}. These works often use isothermal turbulent-box simulations initialized with a double-delta function PDF of pristine gas and enriched gas in some varying spatial distribution, ignoring ongoing enrichment. Initial PDFs of this form were demonstrated some time ago to evolve into a Gaussian distribution at late times \citep{EswaranPope1988}, but it is unclear if these could also be described with a log-normal distribution. However, these works uniformly do not contain the high metallicity power-law tails shown in our work. As suggested by our toy model in Section~\ref{sec:interpretation}, this is due to the lack of ongoing enrichment in these studies.

More detailed models of global galaxy evolution have generally focused on spatial correlations and mixing timescales of initially asymmetric fields, without concern for ongoing enrichment \citep[e.g.][]{deAvillez2002,Petit2015} or focused primarily on the evolution of stellar abundance patterns and metallicity distribution functions \citep[e.g][]{Jeon2017,Hirai2017,Escala2018}, without directly examining the evolution of gas-phase metallicity PDFs.

% The log-normal density PDFs found in ISM simulations have a variance that is proportional to the mach number \citep[e.g.][(maybe more)]{PNJ1997, Passot1998, Fedderath2008, LemasterStone2009}. In our simulations, however,

\subsection{Timescale Dependence of AGB Ejecta}
\label{sec:discussion:metal yields}
We can generally say that metals ejected in AGB winds evolve qualitatively differently than metals produced by SNe in our dwarf galaxy. However, exactly which metals exhibit these differences, and to what degree, is timescale and metallicity dependent. Short timescales ($\lesssim 100$~Myr) only sample enrichment from the most massive AGB stars, while stars on order of a few solar masses only enrich on timescales of a gigayear or longer. In our simulations, which only simulate 500 Myr of evolution, N and Ba are dominated by AGB wind ejecta. C is commonly used to track AGB wind enrichment, but as it originates from lower mass AGB stars, C enrichment operates on gigayear timescales, longer than we follow in this work. Enrichment additionally varies with metallicity. Sr, for example, is only significantly ejected through AGB winds at our metallicity on gigayear timescales, while stellar winds from more massive stars dominate the production of Sr on shorter timescales. At higher metallicity, much more Sr is produced through massive AGB stars, decreasing the timescale over which Sr should exhibit a differential chemical evolution compared to SN-ejected elements (see \cite{Ritter2018} and \cite{Ritter2018b} and references therein).

\begin{figure*}
\centering
\includegraphics[width=0.95\linewidth]{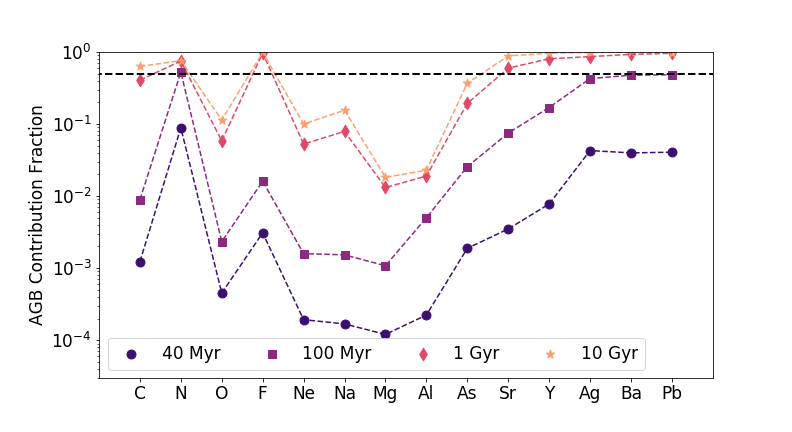}
\caption{The fraction of a given metal ejected through AGB winds at various times for a model of a single-age stellar population at $Z = 10^{-4}$ metal mass fraction, without continuing star formation. We only show a sample of some of the elements dominated most by AGB enrichment at late times. The horizontal line marks a 50\% contribution.}
\label{fig:agb evolution}
\end{figure*}

To illustrate some of these differences, Figure~\ref{fig:agb evolution} gives the fraction of a given metal ejected by AGB winds relative to the total amount of that metal produced in a single-age stellar population with a metal fraction of $Z = 10^{-4}$ at four different times. This model was run using \textsc{SYGMA} \citep{Ritter2018b}. As shown, metal enrichment from AGB winds only begins to dominate for any elements after $\sim$ 100 Myr. This includes N and Ba, which we follow in our simulations, but this model predicts that Ag and Pb will show similar behavior. It takes over a gigayear for C to be dominated by AGB wind ejecta, which is the case for many of the elements shown. F, which shows very little contribution from AGB winds at short timescales, is dominated by enrichment from them on longer timescales. These dependences on metallicity and timescales certainly add complications in generalizing our work and in interpreting observations in the context of the results presented here, but these difference could be leveraged to better understand galactic chemical evolution on multiple, distinct timescales. Clearly this motivates future work covering gigayear timescales to fully understand how metal mass fraction PDFs evolve with time.

\subsection{Impact on Stellar Abundance Patterns}
\label{sec:stellar abundances}
In order to better understand the physics driving stellar abundance patterns and distributions it is important to characterize the chemical evolution of star-forming gas in our simulations. This could be used to help disentangle sources of scatter in observed stellar abundances, including radial/azimuthal abundance gradients and stellar migration, redshift evolution, asymmetric accretion of pristine gas, and the intrinsic scatter in ISM abundances. If the metal mass fractions in star-forming gas in our simulations can also be well fit by log-normal + power-law PDFs, and if we can parameterize their evolution as functions of global galaxy properties, this could be used as a powerful tool for modeling stellar abundance patterns in semi-analytic models. This would be a physically motivated way to account for both the intrinsic spread in stellar abundances due to inhomogeneities in the ISM and variations in mixing for different metal species. We reserve an analysis of these distributions in abundance space in both the gas and stars for future work. However, we briefly discuss the connection to star-forming gas and stellar enrichment below.

\begin{figure}
\centering
\includegraphics[width=0.95\linewidth]{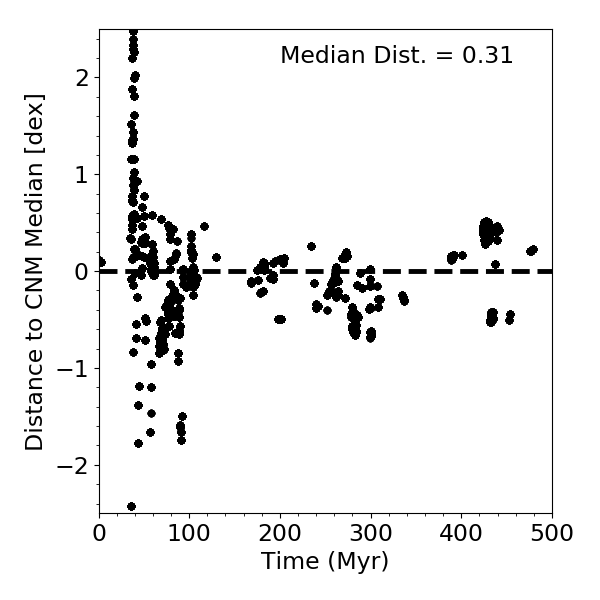}
\caption{The separation (in dex) of each star's oxygen fraction from the median value of the CNM oxygen mass fraction PDF for the time within 1~Myr (our time resolution) of each star's formation time. The median deviation is given in the plot for stars formed after the initial star formation and enrichment period (120~Myr.)}
\label{fig:stars}
\end{figure}

Unfortunately, there is insufficient star-forming gas at any one time in our simulations to construct PDFs of metal mass fractions. However, star-forming gas originates from the CNM, so it is reasonable to expect that this gas, and therefore stars themselves, to have mass fraction PDFs similar to the CNM. To verify this, Figure ~\ref{fig:stars} shows the difference between the oxygen mass fraction of stars and the median mass fraction from the CNM PDF at the time that particle formed. The large scatter at early times ($<$ 120~Myr) is a result of the early enrichment phase, when the initial gas oxygen mass fraction was zero. Stars seem to be sampled evenly around the median of the CNM distribution, with only a slight bias (52\% of stars) towards values below the median. However, for stars formed after the initial phase, the median separation from the CNM median is 0.31~dex, and can reach up to $\sim$~0.5~dex. Though a significant deviation, this is smaller than the typical inner-quartile range (IQR) of the CNM (see Figure~\ref{fig:phase-statistics}). Additionally, if we further subdivide the CNM by density, the metal fraction PDFs narrow and tend towards the median value as a function of increasing density. High density gas, from which star formation occurs, is not biased towards higher metallicity in cold gas.

%KVJ suggested emphasizing that there is much more potential in this work / work like this by examining more dimensions
We emphasize that this study focuses on each chemical dimension independently, as a means to first understand galactic chemical enrichment in the simplest possible framework. However, the cross-correlation of multiple metal fractions and abundance ratios is the most useful in revealing key processes in galactic chemodynamics. A study in this multi-dimensional chemical space is key to understand the relative timescales over which certain enrichment events become important, the number of distinct nucleosynthetic channels that determine observed stellar abundance patterns, and is required for chemical tagging experiments. We will examine this more complicated multi-dimensional space with these numerical methods in future work.

\subsection{Do These Results Apply to More Massive Galaxies?}
\label{sec:massive galaxies}
An important caveat about our work is that these results are derived from simulations of an isolated, low mass, dwarf galaxy whose properties vary dramatically from more massive galaxies like the Milky Way. It is unclear how much these results apply to more massive galaxies with deeper potential wells and higher star formation rates. Feedback-driven galactic outflow properties do vary significantly with halo mass \citep[e.g.][]{MacLowFerrara1999,Muratov2017}, as more massive galaxies more easily retain and re-accrete gas. We expect the results presented in Figure~\ref{fig:species_fractions} to be the most susceptible to the particular star formation history and halo depth of a given galaxy. With gas from individual enrichment events more easily contained, we would expect the retention fractions to be more similar across metal species with increasing halo mass or decreasing star formation rate. We anticipate three potential regimes of metal retention, depending on dark matter halo mass and SFR: i) at very low halo mass, even below that examined here, there is equally poor retention of metal enrichment from both AGB stars and SNe, ii) AGB enrichment is preferentially retained, but SN enrichment is ejected efficiently (as is the case in this study), and iii) both sources are well retained in the galaxy's ISM. If true, this difference in metal retention can be a key observable in verifying our result that the dynamical evolution of metals in the ISM is not uniform.
% AE: rewording given Benoits suggested changes
%     If massive galaxies retain metals equally, we would expect dwarf galaxies to exhibit larger abundance ratios between AGB wind elements and SN enriched elements than stars in more massive galaxies, like the Milky Way.
This difference would be greater at later times, once AGB enrichment becomes significant. 
%     Benoit suggested limiting / eliminating discussion relative to metallicity, since this is a little murky (age-metallicity varies significantly galaxy-to-galaxy). In addition, the elevated signals of Ba abundance ratios I mention have some amount of contention....
%     For example, at metallicities of [Fe/H] $\gtrsim$ 1, Ba is predominantly produced via s-process and ejected in AGB winds \citep{Travaglio1999,Travaglio2004}. At this metallicity, we would expect an elevated [Ba/$\alpha$] at fixed [$\alpha$/H] for low mass dwarfs as compared to the Milky Way. Indeed, this is found in the dSph's of the Milky Way \citep[see ][]{Tolstoy2009} and has been argued to either be the result of variations in s-process enrichment in metal poor environments and/or the result of enrichment timescale differences given the different star formation histories of dwarf galaxies and the Milky Way. In addtion, recent observations find elevationd [Ba/Fe] vs. [Fe/H] in local dwarf galaxies \citep{DugganKirbyAAS}, which could possibly be explained by this effect, but could also point towards significant Ba enrichment from NSMs. Given the lack of definitive explanation, clearly generalizing our results and detailing their observational consequences is a valuable area of future research. 
We would expect dwarf galaxies to exhibit larger abundance ratios between AGB wind elements and SN enriched elements than stars in more massive galaxies, like the Milky Way.

In contrast, we expect the properties of the enrichment PDFs, and the variations between AGB enriched elements and SN enriched elements (as presented in Figures~\ref{fig:log-normal} and ~\ref{fig:phase-statistics}), to be general. We expect metals to exhibit log-normal + power-law distributions in the ISM of all galaxies, with similar trends with enrichment source and gas phase as outlined here. However, the detailed properties of the PDFs (e.g, log-normal width, power-law slope) likely depend non-trivially on global galaxy properties. How these results vary as a function of galaxy properties will be investigated in future work.

\subsection{Implications for Exotic Enrichment Sources}
\label{sec:exotic enrichment}
%   AE: Benoit suggested removing arguments relating to NS-NS merger energetics as I had misred / miscaclulated the energy release values. He said greater emphasise on potential differences of mixing due to rarity, rather than energy.
We have shown that the dynamical evolution of metals depends upon their nucleosynthetic source, focusing here on AGB synthesized elements and SN synthesized elements. However, these differences should also apply to exotic enrichment sources, such as hypernovae, neutron-star neutron-star mergers, and neutron-star black-hole mergers. These sources can have energies that differ significantly from typical SNe, reaching $> 10^{52}$~erg for hypernovae \citep{Nomoto2004}, for example. In addition, 
%mm the rarity of these events, for example,  there are approximately $10^{-5}$ per M$_{\odot}$ of star formation NSM in the Milky Way, could
     these events are rare.  For example, neutron star mergers occur 
     at a rate of approximately $10^{-5}$ per M$_{\odot}^{-1}$ of star
     formation in the Milky Way\citep{Kim2015}.  This rarity could 
significantly influence the typical ISM environments into which they eject their metals. We would therefore expect different mixing behaviors for metals synthesized through these channels as compared to SNe. Based on our results here, for example, we would expect elements from hypernovae to be more well mixed in the ISM of all galaxies, but more readily ejected in dwarf galaxies, as compared to elements from SNe. 
% Benoit suggesed removing this given energy argument removal above
%    Elements from NSMs, a significant source of r-process enrichment, would be less well-mixed in the ISM of all galaxies and more easily retained than SNe elements in dwarf galaxies. 
Depending on their injection energy, neutron star mergers could exhibit different mixing behaviors in the ISM. These differences could provide important signatures for distinguishing individual, exotic enrichment sources from observed stellar abundance patterns in our own Milky Way and in nearby dwarf galaxies. In particular, low mass dwarf galaxies with unusual (compared to Milky Way, solar abundances) r-process enrichment \citep[e.g.][]{Ji2016,Ji2018,Duggan2018}, are valuable for constraining the source of these elements, their frequency, and typical yields. The differential metal evolution presented in our work both opens up an additional avenue by which elements from distinct nucleosynthetic sources may be distinguished in observations and challenges current assumptions used in interpreting these observations.

\subsection{Individual Stars vs. Averaged Yields}
\label{sec:SSP yields}
Typical models for chemical enrichment in galaxy-scale simulations apply some form of IMF-averaged nucleosynthetic yields. Unfortunately, it is beyond the scope of this work to investigate how our results might change when adopting an IMF-averaged yield model. However, to what degree stochastic IMF sampling, mass-dependent nucleosynthetic yields, and the dynamical decoupling of individual enrichment sources play in galactic chemical evolution are valuable questions to address in future research. We speculate that an IMF-averaged enrichment model will only capture the differences between enrichment sources seen here if the model: 1) captures a multi-phase, turbulent ISM, 2) accounts for the energetic differences between yields from different enrichment sources, and 3) accounts for the time-delay between different enrichment sources. Models which average over an entire stellar population and ignore these effects may be unable to reproduce our results.

\section{Conclusions}
\label{sec:conclusions}
We present a detailed analysis of galactic chemodynamics and metal mixing on an element-by-element basis in a low mass dwarf galaxy with hydrodynamics simulations that simultaneously capture multi-channel stellar feedback in detail with a multi-phase ISM. This high resolution simulation, coupled with our star-by-star modeling of stellar nucleosynthetic yields, has allowed us to analyze for the first time how individual metal species couple to the ISM and galactic wind of this galaxy. 

We find that individual metal species do not share the same dynamical evolution, with differences directly related to nucleosynthetic origin (AGB winds, winds from massive stars, or core collapse SNe). This difference is most significant, in our model, between elements ejected predominately by AGB winds and those ejected predominantly by core-collapse SNe. In addition, we find the novel result that the mass fraction PDFs of each metal in the ISM can be described using an analytic piecewise log-normal + power-law PDF. The properties of these PDFs vary with both ISM phase and metal species, again driven mostly by differences in enrichment sources.

We summarize our results as follows:
\begin{itemize}

%\item Elements from lower energy enrichment sources (i.e. AGB winds) are preferentially retained in the ISM of low mass dwarf galaxies as compared to those from higher energy sources (SNe), which are more readily launched in galactic winds. This result is likely sensitive to global galaxy properties, becoming less significant with increasing halo mass. We predict the abundance ratios between AGB wind dominated metals (s-process elements) and SN dominated metals ($\alpha$ elements) to be higher at fixed metallicity in low mass dwarf galaxies than in more massive galaxies, like the Milky Way.

\item Power-law tails on log-normal metal fraction PDFs are a natural consequence of ongoing chemical enrichment, with the power-law slope related to the rate at which mixing dilutes newly enriched gas.

\item Hotter phases have metal fraction PDFs that are more enriched, with significant power-law tails as compared to cold phases, which have a more prominent log-normal component. The lack of significant tails in the cold-phase PDFs indicates that metal mixing occurs rapidly in hotter phases before cooling and/or mixing into denser gas. 

\item Metal outflow in low mass dwarf galaxies depends upon nucleosynthetic site. Metals from lower energy enrichment events (e.g. AGB winds) are preferentially retained in the ISM as compared to those from higher energy events (e.g. SNe). The degree to which this is true likely depends upon global galaxy properties such as star formation rate, dark matter potential well, and gas geometry.

\item Likewise, metals originating in AGB winds are less well mixed in the ISM, with spreads of over 1 dex in cold gas, as compared to metals injected through SNe, with spreads of about 0.5~dex

\item Metal distributions exhibit positive skew, such that the mean metal fraction can be anywhere from 0.1 to 1.0 dex above the median metal fraction. Simple chemical evolution models, which generally follow the mean abundance, and thus can not account for complex metal mixing physics, are likely biased towards higher enrichment values.

\end{itemize}

%mm [you argued above that the wind results would NOT necessarily be general. I'm not sure I like the formulation about entrainment of the ambient ISM, though.] We expect these results to be general for more massive galaxies and galaxy properties. However, it may be the case that metals couple more uniformly to galactic winds in more massive galaxies as wind driving occurs increasingly through entrainment of the ambient ISM rather than the ejection of individual enrichment events. This warrants further examination. But interpreting these results for low mass dwarf galaxies
    Extending these results to low-mass dwarf galaxies in general, we expect that: 1) s-process elements from AGB winds should exhibit larger spreads than $\alpha$ elements released through SNe and 2) these elements should be overabundant in dwarf galaxies at fixed age, as compared to massive galaxies like the Milky Way.
% Removed as-per Benoit's suggestion:
% , and 3) that r-process enrichment from NS-NS mergers should behave similarly to s-process elements from AGB winds, as compared to $\alpha$ elements

\acknowledgments
A.E. is funded by the NSF Graduate Research Fellowship DGE 16-44869. G.L.B. is funded by NSF AST-1312888, NASA NNX15AB20G, and NSF AST-1615955. M.-M.M.L. was partly funded by NASA  grant NNX14AP27G and by NSF grant AST18-15461. B.C. acknowledges support from the ERC Consolidator Grant (Hungary) funding scheme (project RADIOSTAR, G.A. n. 724560) and from the National Science Foundation (USA) under grant No. PHY-1430152 (JINA Center for the Evolution of the Elements). B.W.O was supported in part by NSF grants PHY-1430152 and AST-1514700, by NASA grants NNX12AC98G, and NNX15AP39G, and by Hubble Theory Grant HST-AR-14315.001-A. We gratefully recognize computational resources provided by NSF XSEDE through grant number TGMCA99S024, the NASA High-End Computing Program through the NASA Advanced Supercomputing Division at Ames Research Center, Columbia University, and the Flatiron Institute. This work made significant use of many open source software packages, including \textsc{yt}, \textsc{Enzo}, \textsc{Grackle}, \textsc{Python}, \textsc{IPython}, \textsc{NumPy}, \textsc{SciPy}, \textsc{Matplotlib}, \textsc{HDF5}, \textsc{h5py}, \textsc{Astropy}, \textsc{Cloudy} and \textsc{deepdish}. These are products of collaborative effort by many independent developers from numerous institutions around the world. Their commitment to open science has helped make this work possible. 

\facilities{XSEDE (Stampede, Stampede-2), ADS} % there is no ApJ kewword for this: NASA HECC (Pleiades)}
\software{Numpy, Enzo, yt, astropy}

\bibliographystyle{yahapj}
\bibliography{msbib}

\appendix
\renewcommand\thefigure{\thesection.\arabic{figure}}
\setcounter{figure}{0}

\section{Density PDF}
\label{appendix:density PDF}
\begin{figure}
\centering
\includegraphics[width=0.475\linewidth]{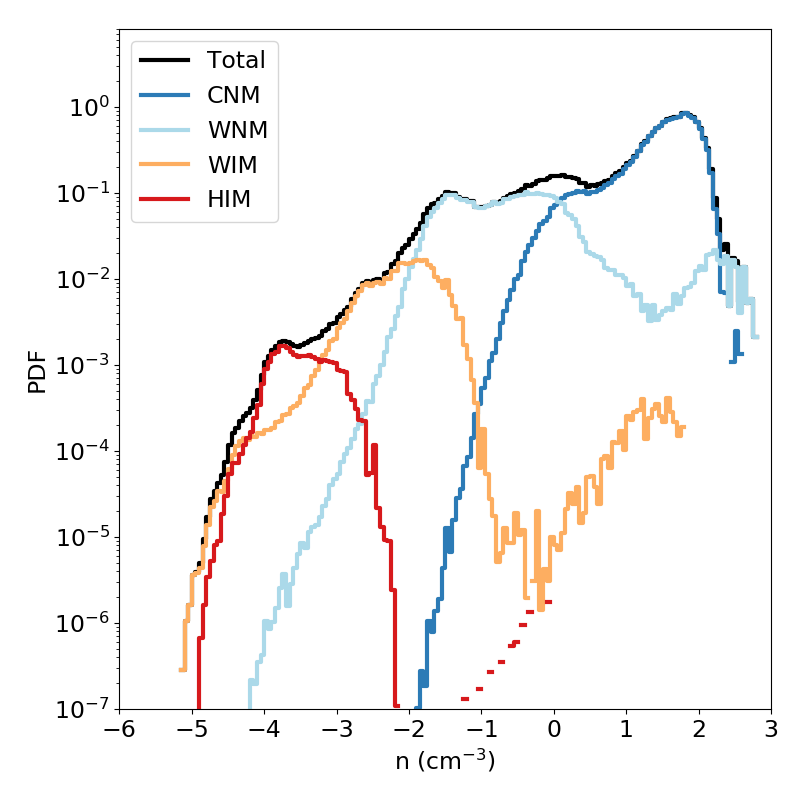}
\includegraphics[width=0.475\linewidth]{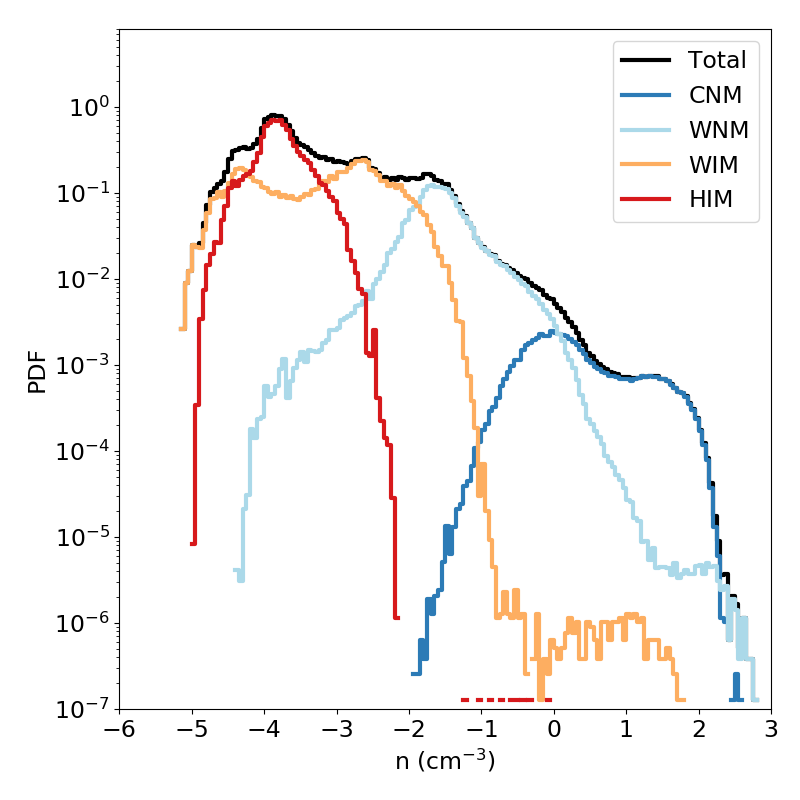}\caption{The mass-weighted (left) and volume weighted (right) density PDFs of our dwarf galaxy at an arbitrarily chosen time of 250~Myr. The total distribution is given in black, sub-divided by the contributions of the individual phases in the ISM.}
\label{fig:density_pdf}
\end{figure}

We show the density PDF in Figure~\ref{fig:density_pdf} to illustrate our result in comparison to comparable works that have computed the density PDF in global galaxy simulations with a multi-phase ISM \citep{Joung2009, Tasker2009, Tasker2011, Tasker2015,HopkinsQuataertMurray2012}. We show the mass-weighted PDF ($dm/Md\log n$) on the left, and the volume-weighted PDF ($dv/Vd\log n$) on the right. As has been demonstrated in previous work, the full density PDF (black) is not well described as a log-normal distribution. The mass-weighted PDF is broad and flat at low densities, with a large tail through to high densities.   The volume weighted PDF is much better described as a multi-component power law. The other phases do also show some log-normality (more for the mass weighted PDFs than the volume weighted PDFs), but all exhibit power-law tails towards higher densities. These deviations from a log-normal may still be the result of grouping together qualitatively different types of ISM gas. 

%mm [you could cut this if you want.]
%It is worth noting that this PDF shows the rather small cold gas 
%     fraction found in these models in comparison to the Milky Way
%      ISM, where roughly half the mass is found in cold gas 
%      \citep{Ferriere2001}.
Finally, \cite{Hopkins2013} suggests a different functional form for describing these PDFs across a range of idealized simulations, but it still may be insufficient to fully describe the density PDF in realistic galaxy simulations. Clearly, we are far from a general understanding of mass-weighted and volume-weighted density PDFs in the case of a turbulent, self-gravitating, multi-phase ISM.

\section{Resolution Comparison}
\begin{figure*}
\centering
\includegraphics[width=0.95\linewidth]{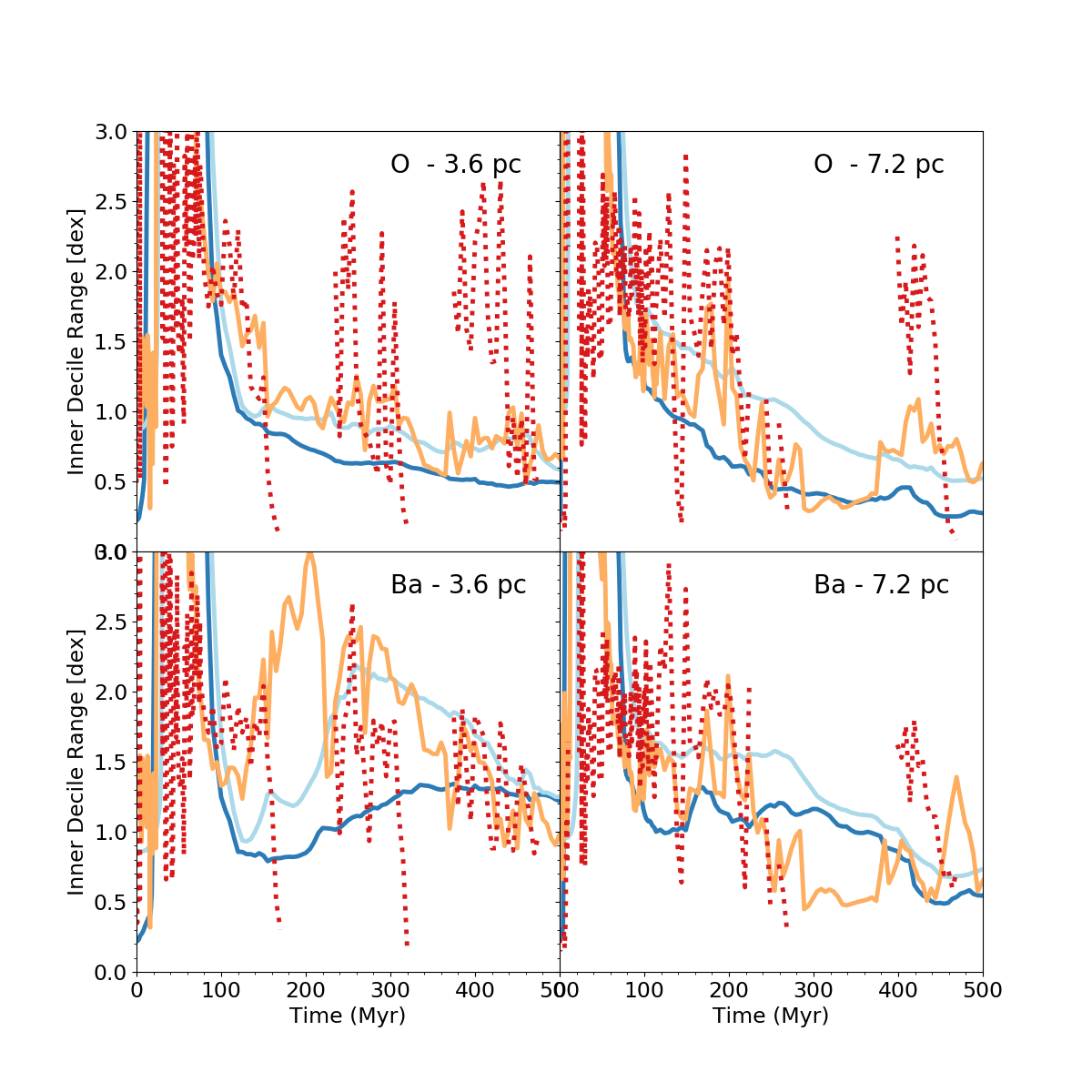}
\caption{Resolution comparison of two lower resolution runs giving the log difference between the 90$^{\rm th}$ decile and 10$^{\rm th}$ decile in dex for the metal PDFS of O and Ba across all phases. Compare to Figure~\ref{fig:phase-statistics}.}
\label{fig:resolution-phase}
\end{figure*}

We perform a resolution test to confirm that the key results of this study are convergent, at least qualitatively. Given the variations in star formation rate, feedback effectiveness, and stochasticity in our model, we do not expect exact numerical convergence in any one quantity. We conduct two lower resolution simulations with a maximum physical resolution of 3.6 pc and 7.2 pc. We refer the reader to Paper I for a previous comparison of these simulations to our fiducial run. In Figure~\ref{fig:resolution-phase} we demonstrate that O and Ba abundances behave qualitatively similar in our lower resolution runs as in our fiducial model. In both lower resolution runs O has a generally tighter distribution that narrows over time, while Ba is much less well mixed, in agreement with our fiducial simulation. The exact numerical values for these spreads are not convergent across simulations, but we do note significant variations in the exact SFH of these lower resolution runs could be driving these differences.

\end{document}